\documentclass[12pt]{article}
\usepackage{amssymb}

\makeatletter
\def\eqnarray{%
 \stepcounter{equation}%
 \let\@currentlabel=\theequation
 \global\@eqnswtrue
 \global\@eqcnt\z@
 \tabskip\@centering
 \let\\=\@eqncr
 $$\halign to \displaywidth\bgroup\@eqnsel\hskip\@centering
 $\displaystyle\tabskip\z@{##}$&\global\@eqcnt\@ne
 \hfil$\displaystyle{{}##{}}$\hfil
 &\global\@eqcnt\tw@$\displaystyle\tabskip\z@{##}$\hfil
 \tabskip\@centering&\llap{##}\tabskip\z@\cr}
\makeatother

\renewcommand{\theequation}{\thesection.\arabic{equation}}

\usepackage{amssymb}

\date{\empty}

\begin{document}

\title{$N=3$\hspace{0.2cm}-\hspace{0.2cm}extended Supersymmetric \\
 Schwarzian and Liouville  Theories} 

\vspace{2cm}

\author{Shogo Aoyama\thanks{Professor emeritus, e-mail: aoyama.shogo@shizuoka.ac.jp}
   \\
       Department of Physics \\
              Shizuoka University \\
                Ohya 836, Shizuoka  \\
                 Japan}

\maketitle

\begin{abstract}
 $N=3$ super-Schwarzian and $N=(3,0)$ super-Liouville theories are formulated by the coadjoint orbit method. 
 We  study the coadjoint orbit dependence of the respective theories,  represented by a superfield $b$. We show that it is renormalized into the $N=3$ super-Schwarzian derivative when the $b$ field takes an appropriate configuration
at the initial point of the orbit. Then the renormalized actions of the respective theories are invariant under   OSp(2$|$3) transformations. If the configuration gets further specified,  
the initial point of the orbit turns out to be  stable under one other kind of  OSp(2$|$3) transformations as well. 
\end{abstract}

\newpage

\section{Introduction}
A discovery of  a duality between the SYK model\cite{SY, K} and the $D=2$ effective gravity raised 
  a vivid interest in the Schwarzian theory. It was considered as playing a role of a mediator between the dual theories.  
Various interesting generalizations of the Schwarzian theory were undertaken. Supersymmetric generalization is one of them. 
$N=1,2$ and $4$ super-Schwarzian theories were discussed in \cite{Super,Mertens, AoHo}. In \cite{AoHo} the 
  $N=4$  theory  was formulated  by  the coadjoint orbit method. The  differential geometry in  the theory then got manifest.  Such a formulation was originally given  for the non-supersymmetric Schwarzian theory in \cite{Wi}. The formulation    can be applied  for the $N=1$ and $2$ theories as well  by simply adjusting the arguments in \cite{AoHo}.  

The  coadjoint orbit method was originated by Alekseev and Shatashvili thirty years ago\cite{Al}.  By the method they studied  the Liouville theory which is one of the $D=2$ effective gravity. Supersymmetric generalization of the arguments  was subsequently undertaken.  The $N=(1,0),(0,1),(1,1) $ and $(2,0)$ super-Liouville theories were discussed in the literature \cite{A2,A2',Del,Ara}. Structural resemblance between these Liouville theories and   
  the  Schwarzian theories is notorious. Further extension  to the 
$N=(4,0)$ super-Liouville  theory  was discussed only recently in \cite{A5}. 

An $N=3$ supersymmetric extension is missing for  the Schwarzian and Liouville theories both. 
The aim of this paper is to study it and  to fill up the gap in the literature.  To this end we need a proper knowledge about 
 the $N=3$ superconformal algebra.   It has been discussed in few occasions in the literature. We begin by giving  a brief summary on the algebra  in Section 2. The reader  familiar with the algebra may skip this section. In Section 3 the coadjoint orbit method is worked out 
to obtain the Kirillov-Kostant two-form $\widehat \Omega_b$ which is closed and invariant under the $N=3$ superconformal diffeomorphism. It is shown  that 
$$
\widehat \Omega_b =dL,\quad\quad\quad i_v\widehat \Omega_b=dH.
$$
Here $L$ is a one-form while $H$ is a zero-form. They  give the $N=3$ super-Liouville and super-Schwarzian theories respectively. (See (3.15).) 
In Section 4 we study  the  orbit dependence of the Kirillov-Kostant two-form $\widehat \Omega_b$, represented  by a superconformal field $b$. It is shown that the $b$ dependence is renormalized into the Kirillov-Kostant two-form $\widehat \Omega_b$ with $b=0$ when   
   the configuration  is chosen appropriately at the initial point of the orbit. Then the renormalized  Kirillov-Kostant two-form gets invariant under OSp(2$|$3) transformation. 
If the configuration   is furthermore specified,   the initial point of the orbit  gets  stable  under one other  kind of OSp(2$|$3) transformation as well.  These symmetries  are    called    
  OSp(2$|$3)$_{target}$ and OSp(2$|$3)$_{diff}$ respectively.  
In Sections 5 and 6 they are examined for 
the   $N=3$ super-Schwarzian and $N=(3,0)$ super-Liouville actions, given in Section 3.   The above twofold OSp(2$|$3) symmetry    of the  renormalized Kirillov-Kostant two-form 
 is recovered for these actions. 
 The  super-Schwarzian action is worked out by expanding superfields in components. It turns out to be remarkably simple. (See (\ref{HHH}).)  We  comment on  its quantization  in the final section. 
As for the super-Liouville theory   the action contains  a non-local term. 
 It can be  hardly put  in a local form in  the supercovariant  formulation with superfields. 
We give it a local expression by expanding  the superfields  in components. The expression is rather complicated. (See (\ref{explicit-gamma}).)
The similar complication  has already appeared in the $N=(4,0)$ theory\cite{A5}, but
not  in the $N=(0,0),(1,0)$ and $(2,0)$  theories.  In this regard we make one more comment on the super-Liouville theories in the final section. 
 Appendices A and B contain helpful formulae for some arguments in the paper. 
 Appendix C is devoted to make clear a subtle difference between  the $N=3$ and 4 super-Schwarzian theories  in arguing 
 the twofold OSp(2$|$3) symmetry. 

\section{$N=3$ superconformal symmetry}

\noindent
{\bf i)}  {\boldmath $N=3$}\ {\bf  superconformal diffeomorphism}

 The $N=3$ superspace is described by  supercoordinates 
$$
(x,\theta_1,\theta_2,\theta_3) \equiv (x,\theta).
$$ 
 Here $x$ is a one-dimensional coordinate  and  $\theta_a,a=1,2,3$ are fermionic coordinates belonging to the O(3) triplet. 
 The supercovariant derivatives are defined by
\begin{eqnarray}
D_{\theta a}={\partial\over\partial\theta_a}+\theta_a\partial_x,  \nonumber
\end{eqnarray}
so as to satisfy 
\begin{eqnarray}
\{D_{\theta a},D_{\theta b}\}=2\delta_{ab}\partial_x.
  \nonumber
\end{eqnarray}
Under a transformation of the supercoordinates
\begin{eqnarray}
x\longrightarrow f(x,\theta_1,\theta_2,\theta_3)\equiv f(x,\theta), \quad\quad \theta_a \longrightarrow \varphi_a(x,\theta_1,\theta_2,\theta_3)\equiv\varphi_a(x,\theta),
  \label{diffeo}
\end{eqnarray}
the supercovariant derivative $D_{\theta a}$   changes  to  $D_{\varphi a}$
\begin{eqnarray}
D_{\theta a}=(D_{\theta a}\varphi^b)D_{\varphi b},
\end{eqnarray}
if the transformation (\ref{diffeo}) satisfies  the  superconformal condition 
\begin{eqnarray}
D_{\theta a}f=\varphi_bD_{\theta a}\varphi_b. 
  \label{cond}
\end{eqnarray}
Then (\ref{diffeo}) is called {\it superconformal diffeomorphism}. 
If a superfield $A(x,\theta)$ transforms by  (\ref{diffeo})  as
\begin{eqnarray}
A(x,\theta)&\longrightarrow& \Delta^wA(f(x,\theta),\varphi(x,\theta)), 
\label{AA}
\end{eqnarray}
with a scaling factor  $\Delta\equiv {1\over 3}(D_{\theta a}\varphi_b)(D_{\theta a}\varphi_b)$, it is called  {\it superconformal field} with weight $w$.\footnote{ 
See Appendix A for an alternative expression of $\Delta$.} 
 Consider   an infinitesimal superconformal diffeomorphism
\begin{eqnarray}
x\longrightarrow f(x,\varphi)=x +\delta x +\cdots, \quad\quad 
\theta_a \longrightarrow  \varphi(x,\varphi)=\theta_a+\delta\theta_a +\cdots.
  \label{infinit}
\end{eqnarray}
It is constrained by (\ref{cond}) as
\begin{eqnarray}
\delta \theta_a={1\over 2}D_{\theta_a}(\delta x+\theta_c\delta \theta_c).
  \label{const-inf}
\end{eqnarray}
Calculating the Lie variation  we find 
(\ref{AA})  in the infinitesimal form\footnote{Hereinafter we do not write the arguments of superfields explicitly if they are simply $(x,\theta)$. } 
\begin{eqnarray}
\delta A&=& ((\delta x+\theta_c\delta\theta_c)\partial_x+\delta\theta_cD_{\theta c}+w\partial_xv)A
\nonumber\\ 
&=&(v\partial_x+{1\over 2}(D_{\theta c} v)D_{\theta c} +w\partial_xv)A.
\label{A}
\end{eqnarray}
Here we used the constraint (\ref{const-inf}) and  defined a parameter of the Lie variation as $v\equiv\delta x+\theta_c\delta\theta_c$. From now on   this Lie variation is denoted by $\delta_v$. It satisfies 
\begin{eqnarray}
[\delta_u,\delta_v]=\delta_{[v, u]},
\nonumber
\end{eqnarray}
with
\begin{eqnarray}
[v,u]\equiv v\partial_x u-v\partial_x v+{1\over 2}D_{\theta_a} vD_{\theta a}u. 
\nonumber
\end{eqnarray}

The $N=3$ super-Schwarzian derivative was given  in \cite{Scou} 
\begin{eqnarray}
{\cal S}(f,\varphi;x,\theta)=2{\varepsilon_{abc} (D_{\theta a}D_{\theta b}\varphi_d) (D_{\theta c}
\varphi_d) \over (D_{\theta e}\varphi_f)(D_{\theta e}\varphi_f)}. 
\label{Schwarzian}
\end{eqnarray}
Consider a further  superconformal diffeomorphism  
$$
f(x,\theta)\longrightarrow F(f(x,\theta),\varphi(x,\theta)),\quad\quad 
\varphi(x,\theta)\longrightarrow \Phi(f(x,\theta),\varphi(x,\theta)),
$$
satisfying the superconformal condition 
$$
D_{\varphi a}F=(\Phi_b D_{\varphi a})\Phi_b. 
$$
It obeys  the composition rule 
\begin{eqnarray}
{\cal S}(F(f,\varphi),\Phi(f,\varphi);x,\theta)=\Delta^{1\over2}{\cal S}(F,\Phi;f,\varphi)+{\cal S}(f,\varphi;x,\theta). 
\label{findiffeo}
\end{eqnarray}
Take the infinitesimal transformation (\ref{infinit}) while fixing  $F$ and $\Phi$.  It then  follows that 
\begin{eqnarray}
\delta_v {\cal S}(F,\Phi;x,\theta)&=&(v\partial_x+{1\over 2}\partial_x v+{1\over 2}(D_{\theta b}v)D_{\theta b})
{\cal S}(F,\Phi;x,\theta)
+ {1\over 6}\varepsilon_{abc}D_{\theta a}D_{\theta b}D_{\theta c}v. 
\label{deltaS}
\end{eqnarray}
We note that  the  transformation law  (\ref{A}) can be consistently extended this way, when $w={1\over 2}$.

Finally we examine the non-supersymmetric part of ${\cal S}(f,\varphi;x,\theta)$.
$f$ and $\varphi_a$ are expanded  in components as  (\ref{expansion}). The non-supersymmetric parts  read as 
\begin{eqnarray}
f(x,\theta)=h+ superpartners , \quad\quad \varphi_a(x,\theta) = \theta_a\rho(x)+superpartners.  
\nonumber
\end{eqnarray}
Put these in (\ref{cond}) and (\ref{Schwarzian}). 
 The superconformal condition  becomes\footnote{This is one of the constraints following from the superconformal condition (\ref{cond}). Other constraints are given in Appendix A. }  
\begin{eqnarray}
\partial_xh=\rho^2 +superpartners.  
\label{nonsusy-const}
\end{eqnarray}
Using this  we find the $N=3$ super-Schwarzian derivative to tend to the   well-known non-supersymmetric limit 
\begin{eqnarray}
{\cal S}(f,\varphi;x,\theta)=\cdots\cdots+\theta_1\theta_2\theta_3\Big\{{\partial_x^3h\over\partial_x h}-{3\over 2}
({\partial_x^2h\over\partial_x h})^2+superpartners\ \Big\}. 
\label{Sexpansion}
\end{eqnarray}

\noindent
{\bf ii)}  {\boldmath $N=3$}\ {\bf  superconformal algebra}

The $N=3$ superconformal transformation  (\ref{A}) may be represented by means of  
 a commutation relation
\begin{eqnarray}
(v\partial_x+{1\over 2}(D_{\theta c}v)D_{\theta c}+w\partial_xv)A=[\int dxd^3 \theta\ v{\cal T},A].
\label{operator}
\end{eqnarray}
 Here $\cal T$ is the generator of the  transformation, while  $A$ is regarded as  an operator representing  the superconformal superfield defined by (\ref{AA}).  
 Note that $\cal T$   is a fermionic operator with  weight ${1\over 2}$.  
When  $A$ is taken to be  $\cal T$,  the transformation (\ref{operator}) may become  anomalous  like  that of the Schwarzian derivative, i.e., 
\begin{eqnarray}
(v\partial_x+{1\over 2}(D_\theta v)D_\theta +{1\over 2}\partial_xv){\cal T}+{c\over 6}\varepsilon_{abc}D_{\theta a}D_{\theta b}D_{\theta c}v=[\int dxd^3 \theta\ v{\cal T},{\cal T}],
\label{operator'}
\end{eqnarray}
with a constant $c$. 
It   can be written  in  the standard form of the $N=3$ superconformal algebra
 by taking the Fourier expansion\cite{Ademolo}. To see this we begin by taking expansion of  
 $v$ and   $\cal T$  in $\theta$ as   
\begin{eqnarray}
v&=&\alpha(x)+\theta_c\beta_c(x)+{1\over 2}\varepsilon_{abc}\theta_a\theta_bt_c(x)+
{1\over 3!}\epsilon_{abc}\theta_a\theta_b\theta_c\beta(x),
 \label{v}\\
{\cal T}&=& F(x)+\theta_aT_a(x)+{1\over 2}\epsilon_{abc}\theta_a\theta_bF_c(x)+{1\over 3!}\epsilon_{abc}\theta_a\theta_b\theta_cT(x).
\label{T}
\end{eqnarray}
The resulting components are expanded in  Fourier modes.  
They take the forms 
\begin{eqnarray}
\alpha(x)&=&\sum_{n\in {\bf Z}} e^{in x}\alpha_{ n}, \quad\quad 
\beta_a(x)=\sum_{r\in {\bf Z}+\nu}  e^{i rx}\beta_{a r},  \nonumber\\
t_a(x)&=&\sum_{n\in {\bf Z}}  e^{inx}t_{an}, \quad\quad
\beta(x)=\sum_{r\in {\bf Z}+\nu}  e^{rx}\beta_{ r},  
\label{modeexp}
\end{eqnarray}
and 
\begin{eqnarray}
 F(x)&=&\sum_{r\in {\bf Z}+\nu}e^{irx}F_r, \quad\quad T_a(x)=\sum_{n\in {\bf Z}}e^{inx}T_{an},  \nonumber\\
F_c(x)&=& \sum_{r\in {\bf Z}+\nu}e^{irx} F_{cr},\quad\quad 
T(x)=\sum_{n\in {\bf Z}}e^{inx}L_n,
\nonumber
\end{eqnarray}
in which  $\nu=0$ for the R sector or ${1\over 2}$ for the NS sector. 
By these Fourier expansions we find   
\begin{eqnarray}
\int dx d^3\theta\ v{\cal T}\propto \sum_{n\in {\bf Z}}( \alpha_{-n} L_n+t_{a-n}T_{an}) +\sum_{r\in {\bf Z}+\nu}(\beta_{a-r} F_{ar}
+\beta_{-r}F_r). 
\nonumber
\end{eqnarray} 
 The r.h.s. of (\ref{operator'}) can be similarly calculated. Comparing the Fourier modes of  both sides
 we then find  the $N=3$  superconformal algebra\cite{Miki}
\begin{eqnarray}
[L_m,L_n]&=&(m-n)L_{m+n}+cm(m^2-1)\delta_{m+n,0}, \nonumber\\
\ [T_{a\hspace{0.05cm} m},T_{b\hspace{0.05cm} n}]&=&i\varepsilon_{abc}T_{c\hspace{0.05cm} m+n}+4cm\delta_{ab}\delta_{m+n,0},  
\nonumber\\
\ [L_m,T_{a\hspace{0.05cm}n}]&=& -nT_{a\hspace{0.05cm}m+n}, \nonumber\\
\ [L_m,F_{a\hspace{0.05cm}r}]&=&({m\over 2}-r)F_{a\hspace{0.05cm}m+r},\nonumber\\
\ [T_{a\hspace{0.05cm}m},F_{b\hspace{0.05cm}r}]&=&-i\varepsilon_{abc}F_{c\hspace{0.05cm}m+r}+m\delta_{ab}F_{m+r}, \nonumber\\
\ \{F_{a\hspace{0.05cm}r},F_{b\hspace{0.05cm}s}\}&=& -2\delta_{ab}L_{r+s}+(r-s)i\varepsilon_{abc}T_{c\hspace{0.05cm}r+s}-4c(r^2-{1\over 4})\delta_{ab}\delta_{r+s,0},   \nonumber\\
\ [L_m,F_{\hspace{0.05cm}r}]&=& -({m\over 2}+r)F_{m+r}, \nonumber\\
\ [F_r,T_{a\hspace{0.05cm}m}]&=&0, \nonumber\\
\ \{F_{a\hspace{0.05cm}r},F_s\}&=&-T_{a\hspace{0.05cm}r+s}, \nonumber\\
\{F_r,F_s\}&=& -4c\delta_{r+s,0},
\nonumber
\end{eqnarray}
after scaling  and shifting as\footnote{Further scaling  $F_a\rightarrow i F_a$ and $  F\rightarrow i F$ are needed to get   the $N=3$ superconformal algebra with the normalization   in \cite{Miki,Yama}.
See also Footnote 5.}   
$$
F_a\rightarrow {1\over 2\sqrt i}F_a, \ F\rightarrow {1\over 2\sqrt i}F, \ T_a\rightarrow {1\over 2}T_a, \ L_0\rightarrow L_0-{c\over 2}. 
$$
One can eliminate $F_{r}$ and redefine the remaining generators  to find
the non-Lie algebraic O($3$) superconformal algebra, in the sense that the anti-commutator
 $\{F_{a\hspace{0.05cm}r},F_{b\hspace{0.05cm}s}\}$ contains a  term quadratic in $T_{an}$\cite{Kniz,Ber}.

\noindent
{\bf iii)} {\bf The superalgebra {\boldmath $\frak {osp}(2|3)$}}

 The following twelve zero-mode generators in the above $N=3$ superconformal algebra 
\begin{eqnarray}
 L_{\pm n}, L_0,F_{a\hspace{0.05cm}\pm{1\over 2}n},T_{a\hspace{0.05cm} 0}(=\varepsilon_{abc}T_{[bc]\hspace{0.05cm}0}), \quad\quad n \in {\bf N}_{\ne 0},  \label{subalgebra}
\end{eqnarray}
 form the superalgebra $\frak {osp}(2|3)$  
 By  scaling them  as 
\begin{eqnarray}
&\ & \overline L=nL_{n},\quad  L=nL_{-n},\quad L^0=nL_0, \nonumber\\
&\ &\hspace{-0.8cm} \overline F_a=\sqrt nF_{a\hspace{0.05cm}{1\over 2}n}, \quad  
F_a=\sqrt nF_{a\hspace{0.05cm}-{1\over 2}n},\quad T_a=T_{a\hspace{0.05cm}0},
\nonumber
\end{eqnarray}
it is given  by
\begin{eqnarray}
&\ & [T_a,T_b]=i\varepsilon_{abc}T_c, \quad [T_a,L]=0,\quad[T_a,\overline L]=0,\quad[T_a,L^0]=0,   \nonumber\\
&\ & \hspace{1.5cm}[\overline L,L]=2L^0, \quad [L,L^0]=-L, \quad [\overline L,L^0]=\overline L, \nonumber\\
&\ &\hspace{2.2cm} [F_a, L^0]=-{1\over 2}F_a,\quad [{\overline F}_a,L^0]={1\over 2}{\overline F}_a,\nonumber\\
&\ & [F_a,L]=0,\quad [{\overline F}_a,L]= F_a,  \quad
 [F_a,\overline L]=-{\overline F}_a, \quad [{\overline F}_a,\overline L]=0, \nonumber\\
&\ &\hspace{2cm} [F_a,T_b]=i\varepsilon_{abc}F_a,\quad [{\overline F}_a,T_b]=i\varepsilon_{abc}{\overline F}_c, \nonumber\\
&\ &\hspace{-1cm} \{F_a,F_b\}=-2\delta_{ab}L, \quad \{{\overline F}_a,{\overline F}_b\}=-2\delta_{ab}\overline L, \quad \{{\overline F}_a,F_b\}=-2\delta_{ab}L^0-i\varepsilon_{abc}T_c. 
\label{superalgebra}
\end{eqnarray}

\noindent
{\bf iv) Non-linear realization of {\boldmath $\frak {osp}(2|3)$}} 

 Let us write the above superalgebra $\frak {osp} (2|3)$ in a collective form 
$$
  \delta_\epsilon T^B\equiv [\epsilon_AT^A, T^B]=\epsilon_Af^{AB}_{\ \ \ C}T^C,
$$
with
$
T^A=\{L,{L^0},{\overline L},{F_a}, {\overline F_a}\}
$.  
Here $\epsilon_A$ are infinitesimal parameters, which are graded correspondingly to the grading of $T^A$. This commutation relation may be 
realized by  non-linear transformations  of the  $N=3$ superconformal diffeomorphisms $f$ and $\varphi_a$
\begin{eqnarray}
\delta_\epsilon f 
= \epsilon_AR^A(f,\varphi),   \quad\quad 
\delta_\epsilon \varphi_a  
= \epsilon_AR^A_{\ a}(f,\varphi).    \label{variations}
\end{eqnarray}
Here $R^A(f,\varphi)$ and $R^A_{\ a}(f,\varphi)$ are the Killing vectors realizing  the superalgebra $\frak {osp} (2|3)$  as
$$
 [\delta_\epsilon,\delta_{\epsilon'}] =\delta_{[\epsilon,\epsilon']},
 $$
 with 
 $$
 [\epsilon,\epsilon']=\epsilon_A\epsilon'_B f^{AB}_{\ \ \ C}T^C.
$$
Explicit expressions  of the Killing vectors are found by 
considering the coset space ${\rm OSp}(2|3)/ \hat{\rm H}$ with $\hat{\rm H}$ the subgroup generated by $\overline L,\overline F_a,L_0,T_a$. Write a coset element 
$$
e^{fL+\varphi_aF_a}
$$
with the $N=3$ superconformal diffeomorphisms $f$ and $\varphi_a$. Multiply  $e^{\epsilon_AT^A}$ from the left.  Then we follow the procedure elaborated  in \cite{AoHo,Aolast, Ho} to find the non-linear transformations (\ref{variations}) 
as 
\begin{eqnarray}
\epsilon_AR^A(f,\varphi) &=& \epsilon_L+\epsilon_{L^0}f+\epsilon_{\overline L}f^2-\epsilon_{F\hspace{0.05cm}c}\varphi_c- \epsilon_{\overline F\hspace{0.05cm}c}\varphi_c f,   \nonumber\\
\epsilon_AR^A_{\ a}(f,\varphi)&=& \epsilon_{F\hspace{0.05cm}a}+\epsilon_{\overline F\hspace{0.05cm}a}f
-\epsilon_{\overline F\hspace{0.05cm}c}\varphi_c\varphi_a+{1\over 2}\epsilon_{L^0}\varphi_a+i\epsilon_{abc}\epsilon_b\varphi_c+\epsilon_{\overline L}\varphi_af .
\label{Killing}
\end{eqnarray} 
 It is important that they satisfy the superconformal condition (\ref{cond}) as 
\begin{eqnarray}
D_{\theta a}\delta_\epsilon f=\delta_\epsilon\varphi_aD_{\theta b}\varphi^b+
\varphi_aD_{\theta b}\delta_\epsilon\varphi^b.
\nonumber
\end{eqnarray}
It is also important that 
the $N=3$ super-Schwarzian derivative (\ref{Schwarzian}) is invariant under these non-linear transformations. 
To see this it is sufficient to show that the numerator and the denominator scale respectively as
\begin{eqnarray}
 \delta_\epsilon(\varepsilon_{abc}D_{\theta a}D_{\theta b}\varphi_d D_{\theta c}\varphi_d )
&=&(\epsilon_{L^0}+2\epsilon_{\overline L}z+2\epsilon_{\overline F\hspace{0.05cm} g}\varphi_g)
(\varepsilon_{abc}D_{\theta a}D_{\theta b}\varphi_d D_{\theta c}\varphi_d ),
 \nonumber\\
\delta_\epsilon (D_{\theta e}\varphi_f D_{\theta e}\varphi_f )&=&(\epsilon_{L^0}+2\epsilon_{\overline L}z+2\epsilon_{\overline F\hspace{0.05cm} g}\varphi_g)D_{\theta e}\varphi_f D_{\theta e}\varphi_f, 
\label{SchwarzInv}
\end{eqnarray}
under the non-linear transformations (\ref{variations}).

The whole arguments in this section can be similarly done for the $N=0,1,2$ superconformal symmetries. The zero-mode generators of the resulting $N$-extended superconformal algebra  form the superalgebra $\frak {osp}(2|N)$ with $a,b,c =1,\cdots,N$. 
But the argument for the $N=4$ superconformal algebra goes differently.  
The superalgebra $\frak {osp}(2|4)$ contains $\frak {psu}(1,1|2)$. It is this    
subalgebra which characterized  the $N=4$-extended supersymmetric  Schwarzian and Liouville theories\cite{AoHo,A5}.\footnote{The subalgebra $\frak {osp}(2|3)$ has been normalized in Section 2 consistently with the normalization of these   subalgebrae given in \cite{A2,A3,AoHo,A5}. } 
 The superalgebra $\frak {psu}(1,1|2)$ was also non-linearly realized by  the Killing vectors on an appropriate  coset space. 
But the $N=4$ super-Schwarzian derivative is no longer invariant under those non-linear transformations. 
 Instead it transforms as    
\begin{eqnarray}
\delta_\epsilon{\cal S}=\epsilon_{L^0}+2f\epsilon_{\overline L}+2\varphi_c\epsilon_{\overline F}^{\ c}+2\varphi^c\epsilon_{\overline F c}, 
\label{N4}
\end{eqnarray}
with  the  notation adapted to the $N=4$ superconformal algebra\cite{AoHo,A5}.

\section{The Kirillov-Kostant two-form}

\setcounter{equation}{0}

  The adjoint  and coadjoint actions in the $N=3$ superconformal algebra 
 are respectively given  by a generator $v $ as
\begin{eqnarray}
\delta_v u&=&v\partial_xu-(\partial_xv)u+{1\over 2}(D_{\theta c}v)D_{\theta c}u\equiv[v,u],
 \nonumber\\
\delta_v \widehat b&=&v\partial_x \widehat b+{1\over 2}(\partial_x v)\widehat b+{1\over 2}(D_{\theta c}v)D_{\theta c}\widehat b +cD_{\theta 1}D_{\theta 2}D_{\theta 3}v. 
\label{coad}
\end{eqnarray}  
Here $u,v$ are the $N=3$ superconformal fields with $w=-1$ in the $N=3$ superspace, discussed in Section 2. But $\widehat b$ is the one with $w={1\over 2}$ which transforms anomalously.
  It is  assumed to be fermionic. 
 On top of $u,v,\widehat b$ we  consider  also $f$ and $\varphi_a$, which represented 
   the $N=3$ superconformal diffeomorphisms  in Section 2. Let them to be  $N=3$ superconformal fields with  $ w=0$, i.e.,  
\begin{eqnarray}
\delta_v f&=& v\partial_x f+{1\over 2}(D_{\theta c}v)D_{\theta c}f,
  \nonumber\\
\delta_v \varphi_a&=&  v\partial_x\varphi_a +{1\over 2}(D_{\theta c}v)D_{\theta c}\varphi_a. \nonumber
\end{eqnarray}
Keep in mind that they satisfy the superconformal condition (\ref{cond}). 
 Correspondingly we define the exterior derivative as
\begin{eqnarray}
d u&=&y\partial_xu-(\partial_xy)u+{1\over 2}(D_{\theta c}y)D_{\theta c}u\equiv[y,u],
 \nonumber\\
d \widehat b&=&y\partial_x \widehat b+{1\over 2}(\partial_x y)\widehat  b+{1\over 2}(D_{\theta c}y)D_{\theta c}\widehat b +cD_{\theta 1}D_{\theta 2}D_{\theta 3}y,
\label{d-coad}
\end{eqnarray}  
 and
\begin{eqnarray}
d f&=& y\partial_x f+{1\over 2}(D_{\theta c}y)D_{\theta c}f,
  \nonumber\\
d \varphi_a&=&  y\partial_x\varphi_a +{1\over 2}(D_{\theta c}y)D_{\theta c}\varphi_a.  \label{eq-for y}
\end{eqnarray}
Here $y$ is  a one-form, while $u,\widehat b, f, \varphi_a$ are  zero-forms.  The two derivatives $\delta_v$ and $d$ are related by 
exchanging $v$ and $y$. We introduce a third derivative $i_v$  which replaces $y$ by  $v$, called {\it skew-derivative}.  
 It is assumed to give zero 
 operating  on  zero-forms. Then they   satisfies the  well-known identity in the differential geometry\cite{KN,W}  
\begin{eqnarray}
\delta_v=di_v+i_vd. 
\label{Identity}
\end{eqnarray} 
 Solving the equations in (\ref{eq-for y}) 
 for $y$ with the superconformal condition (\ref{cond}) 
we get 
$$
y={df+\varphi_cd\varphi_c\over \partial_xf+\varphi_c\partial_x\varphi_c}. 
$$
Then it is easy to show\footnote{To show (\ref{dy'})  the formula (\ref{useful}) in Appendix is useful. 
 To show (\ref{deltay'}) we do not need calculation  noting   that the nominator of $y$ has weight  $w =0$, while  the denominator $w=1$. }  
\begin{eqnarray}
dy&=&y\partial_x y +{1\over 4}(D_{\theta c}y)(D_{\theta c}y)={1\over 2}[y,y],
 \label{dy'}\\
 \delta_vy&=& v\partial_xy-(\partial_xv)y+{1\over 2}(D_{\theta c}v)D_{\theta c}y\equiv[v,y]={1\over 2}i_v[y,y].
 \label{deltay'}
\end{eqnarray}
It is instructive to see that the identity (\ref{Identity}) holds for these relations. 

The Kirillov-Kostant two-form  $\Omega$ is defined by  
$$
\Omega =\int dx d^3\theta\omega \equiv {1\over 2}\int dx d^3\theta\widehat b[y,y]. 
$$
A smart way to calculate $d\Omega $ and $\delta_v\Omega$ is to use the  identity relations for the two derivatives 
\begin{eqnarray}
d ( \widehat bu) &=& (d \widehat b)u+\widehat b[y,u]=\partial_x(y\widehat bu)-D_{\theta_c}
\{{1\over 2}(D_{\theta_c}y)\widehat bu \} + c(D_{\theta 1}D_{\theta 2}D_{\theta 3}
 y) u, \nonumber\\
\delta_v ( \widehat bu) &=& (\delta_v \widehat b)u+\widehat b[v,u]=\partial_x(v\widehat bu)-D_{\theta_c}
\{{1\over 2}(D_{\theta_c}v)\widehat bu \} + c(D_{\theta 1}D_{\theta 2}D_{\theta 3}
 v) u.  
\nonumber 
\end{eqnarray} 
They hold even with  $u$ replaced by $[y,y]$, since both have $w=-1$. 
By means of these identity relations 
 we calculate $d\omega$ and $\delta_v\omega$ as
\begin{eqnarray}
d\omega&=&{1\over 2}(d\widehat b)[y,y]  \nonumber\\
&=& {1\over 2}\partial_x(y \widehat b[y,y])- {1\over 2}D_{\theta_c}
\{{1\over 2}(D_{\theta_c}y)\widehat b[y,y] \} +  {1\over 2}c(D_{\theta 1}D_{\theta 2}D_{\theta 3}
 y) [y,y],
\label{d-omega}\\
\delta_v\omega&=& {1\over 2}(\delta_v\widehat  b)[y,y] - {1\over 2}\widehat  b[[y,y],v]={1\over 2}(\delta_v\widehat  b)[y,y] +
{1\over 2}\widehat b\delta_v([y,y])\nonumber\\
&=&  {1\over 2}\partial_x(v\widehat b[y,y])-{1\over 2}D_{\theta_c}
\{{1\over 2}(D_{\theta_c}y)\widehat b[y,y] \} +{1\over 2}c(D_{\theta 1}D_{\theta 2}D_{\theta 3}
 v) [y,y].
\end{eqnarray} 
Here use was made of the Jacobi identity for the commutator. Then we find 
\begin{eqnarray}
d\Omega &=&  {c\over 2}\int dxd^3\theta (D_{\theta 1}D_{\theta 2}D_{\theta 3}y)[y,y]={c\over 2}d\int dxd^3\theta y(D_{\theta 1}D_{\theta 2}D_{\theta 3})y, \nonumber\\
\delta_v\Omega &=& {c\over 2}\int dx d^3\theta (D_{\theta 1}D_{\theta 2}D_{\theta 3}v)[y,y] =c d \int dxd^3\theta  (D_{\theta 1}D_{\theta 2}D_{\theta 3}v)y,  \nonumber\\
i_v\Omega&=&\int dxd^3\theta \widehat b[v,y],
\label{modifications}
\end{eqnarray}
by using   $dv=0$.\footnote{Operation of the exterior derivative $d$ is considered in an extra dimensional space, so that it goes through $\int dxd^3\theta$. }  The consistency of these calculations can be checked by  the identity (\ref{Identity}). 
 $\Omega$ is not invariant under the superconformal transformation.
Hence let us modify it by a counter term 
\begin{eqnarray}
\widehat\Omega= \Omega + {c\over 2}\int dxd^3\theta yD_{\theta 1}D_{\theta 2}D_{\theta 3}y.
\label{Omega-mod}
\end{eqnarray}
The key point for the modification   is  that 
  the counter term  is closed  
\begin{eqnarray}
d \int dxd^3\theta y D_{\theta 1}D_{\theta 2}D_{\theta 3}y=0, 
\label{Omega-closure}
\end{eqnarray}
and  satisfies 
\begin{eqnarray}
\delta_v \int dxd^3\theta y D_{\theta 1}D_{\theta 2}D_{\theta 3}y=d\int dxd^3\theta
(2vD_{\theta 1}D_{\theta 2}D_{\theta 3}y). 
\label{delta-counter}
\end{eqnarray}
They are both shown in Appendix B by direct calculations. But it is worth noting that 
the latter formula follows from the former by means of the identity (\ref{Identity}). 
From these  formulae  it follows  that 
\begin{eqnarray}
d\widehat\Omega= 0,\quad\quad \delta_v\widehat\Omega=0,  
\quad\quad di_v\widehat \Omega=0.  
\label{closure''}
\end{eqnarray}
Due to (\ref{Omega-closure}) there exists a one-form $\gamma$ such as 
\begin{eqnarray}
 \int dxd^3\theta y D_{\theta 1}D_{\theta 2}D_{\theta 3}y=d\int dxd^3\theta \gamma. 
\label{gamma'}
\end{eqnarray}
Then we find from (\ref{closure''}) that 
\begin{eqnarray}
\widehat\Omega= d L,\quad\quad i_v\widehat\Omega=dH, 
\label{final formula}
\end{eqnarray}
in which 
\begin{eqnarray}
L=-\int dxd^3\theta (\widehat by+{c\over 2}\gamma), \quad\quad 
H=\int dxd^3\theta \widehat b v. 
\label{LH-def}
\end{eqnarray}

The arguments so far done  work for any $N=3$ superconformal field $\widehat b$ as long as it transforms as
(\ref{coad}).  
We consider one other  superconformal field $b$ obeying the same transformation law  but with $c=0$.  The whole arguments may be repeated  for  a sum $\widehat b+ b$, since it  still transforms as (\ref{coad}). 
Let $\widehat b$ to be given by $\widehat b=c{\cal S}(f,\varphi;x,\theta)$ with   the $N=3$ super-Schwarzian derivative  (\ref{Schwarzian}). Its transformation law 
  reads from (\ref{deltaS}) with a suitable  change of the notation  as 
\begin{eqnarray}
\delta_v \Big(c{\cal S}(f,\varphi;x,\theta)\Big) &=& v\partial_x \Big(c{\cal S}(f,\varphi;x,\theta)\Big)
+{1\over 2}(\partial_xv) \Big(c{\cal S}(f,\varphi;x,\theta)\Big) \nonumber\\
&+&{1\over 2}(D_{\theta c}v)D_{\theta c}\Big(c{\cal S}(f,\varphi;x,\theta)\Big) 
  +cD_{\theta 1}D_{\theta 2}D_{\theta 3}v,
 \label{Schwarz2}
\end{eqnarray}
As for $b$  we identify  it  with $\Delta^{1\over 2}b(f,\varphi)$. It transforms as  
\begin{eqnarray}
\hspace{-0.5cm}\delta_v \Big(\Delta^{1\over 2}b(f,\varphi)\Big)=v\partial_x\Big(\Delta^{1\over 2}b(f,\varphi)\Big) +{1\over 2}(\partial_xv) 
 \Big(\Delta^{1\over 2}b(f,\varphi)\Big)+{1\over 2}(D_{\theta c}v)D_{\theta c}\Big(\Delta^{1\over 2}b(f,\varphi)\Big), \nonumber\\
 \label{b-transf}
\end{eqnarray}
as can be checked  from the transformation law (\ref{A}) for $b(x,\theta)$ by a direct calculation.
Then 
the sum  $\widehat b+b$ can be obtained from $b(x,\theta)$ as a flow  
\begin{eqnarray}
b(x,\theta) \longrightarrow  \Delta^{1\over 2}b(f,\varphi)+c{\cal S}(f,\varphi;x,\theta), 
\label{map}
\end{eqnarray}
under the $N=3$ superconformal diffeomorpism 
$$
x \longrightarrow f(x,\theta) \quad\quad \theta_a\longrightarrow\varphi_a(x,\theta). 
$$
We think of an orbit  connecting both ends of the transformation (\ref{map}). 
It is a line which starts at $b$ in a space of all  superconformal fields transforming  as $\widehat b$,  called {\it coadjoint orbit}   ${\cal O}_b$.  
It is important to note that
$$
\lim_{f=x,\varphi=\theta} \delta_v(\widehat b+b) \ne \delta_v\Big(\lim_{f=x,\varphi=\theta} (\widehat b+b)\Big)=v\partial_xb+{1\over 2}(\partial_xv)b+{1\over 2}(D_{\theta c}v)D_{\theta c}b, 
$$
at the initial point of the orbit. 
With $\widehat b$ replaced by $\widehat b+b$
we can define the Kirillov-Kostant two-form (\ref{Omega-mod})  on the coadjoint orbit ${\cal O}_b$ as
\begin{eqnarray}
\widehat\Omega_b=\int dxd^3\theta\Big\{\Big(\Delta^{1\over 2}b(f,\varphi)+c {\cal S}(f,\varphi;x,\theta)\Big)[y,y]+{c\over 2}y D_{\theta 1}D_{\theta 2}D_{\theta 3}y\Big\}. 
\label{KKorbit}
\end{eqnarray} 
All the formulae in (\ref{closure''})  hold for $\widehat\Omega_b$. So do those  in  (\ref{final formula}) with 
\begin{eqnarray}
L&=&-\int dxd^3\theta \Big(\Big\{\Delta^{1\over 2}b(f,\varphi)+c {\cal S}(f,\varphi;x,\theta)\Big\}y+{c\over 2}\gamma\Big),\nonumber\\
H&=&\int dxd^3\theta \Big(\Delta^{1\over 2}b(f,\varphi)+c {\cal S}(f,\varphi;x,\theta)\Big)v. 
\nonumber
\end{eqnarray}
In the next sections they are  respectively taken   to be  the actions of the $N=3$ super-Liouville and  super-Schwarzian theories. (\ref{closure''}) is the most salient feature of the Kirillov-Kostant two-form in the coadjoint orbit method.

Finally we shall give a concrete expression for the the assumed one-form $\gamma$
 in (\ref{gamma'}).  
 However it is hard to find a local expression of $\gamma$ in  the supercovariant formulation, i.e.,  in terms of  superconformal fields.\footnote{ Such a  local expression  of $\gamma$ could not be found  for the $N=(4.,0)$ case either\cite{AoHo}. But it
 can be  for $N\le 2$ as commented  in the end of  this paper.} Therefore we look for it in the component formulation.  To this end the author has done rather massive calculations, using the expansion formulae of $f$ and $\varphi_a$ in Appendix A.  
The details of the calculations were exposed in \cite{Note}. We quote only the result
\begin{eqnarray}
&\ &\hspace{0.5cm}6\gamma =-6\Big\{\log\rho^2\partial_xd(\log\rho^2)+\partial_x({1\over\rho^2})\partial_x({1\over\rho^2})
dh\rho^2)\Big\}  \nonumber\\
&\ &\hspace{2.5cm}+\Big\{\hspace{-0.1cm}-6{\partial_x\rho^2\over\rho^2}{\partial_x\rho^2\over\rho^2}{dh\over\rho^2}
\Big({\eta\over\rho}\cdot{\partial_x\eta\over\rho}\Big)
+48{dh\over\rho^2}{\partial_x\rho\over\rho}\Big({\eta\over\rho}\cdot\partial_x^2({\eta\over\rho})\Big)
\nonumber\\
&\ & \hspace{3.5cm} +24{dh\over\rho^2}\Big(\partial_x({\eta\over\rho})\cdot\partial_x^2({\eta\over\rho})\Big)\Big\}
\nonumber\\
&\ &\hspace{2.5cm}+\Big\{\hspace{0cm}24{d\rho^2\over\rho^2}\Big(\partial_x^2({\eta\over\rho})\cdot {\eta\over\rho}\Big)
-24\Big(\partial_x({\eta\over\rho})\cdot\partial_xd({\eta\over\rho})\Big)\Big\}
\nonumber\\
&\ &\hspace{2.5cm}+\Big\{\hspace{-0.1cm}-12{\partial_x\rho^2\over\rho^2}{d\rho2\over\rho2}
\Big({\eta\over\rho}\cdot\partial({\eta\over\rho})\Big)^2
+24{\partial_x\rho\over\rho}\Big({\eta\over\rho}\cdot d({\eta\over\rho})
\Big({\eta\over\rho}
\cdot \partial_x^2({d\rho\over\rho})\Big)
\nonumber\\
&\ &\hspace{3.5cm}
+6\Big(\partial_x({\eta\over\rho})\cdot {\eta\over\rho})\Big)^2
\partial_x({d\rho^2\over\rho^2})
\nonumber\\
&\ &\hspace{4.5cm}
-24\partial_x({\partial_x\rho\over\rho})
\Big(d({\eta\over\rho})
\cdot{\eta\over\rho}\Big)\Big(\partial_x({\eta\over\rho})\cdot
{\eta\over\rho}\Big)
\nonumber\\ 
&\ &\hspace{4.5cm}-24({\partial_x\rho\over\rho})^2\Big(d({\eta\over\rho})
\cdot{\eta\over\rho}\Big)\Big)\Big({\eta\over\rho}\cdot \partial_x({\eta\over\rho})\Big)\Big\}
\nonumber\\
&\ &\hspace{2.5cm}+\Big\{\hspace{-0.1cm}-24\epsilon_{lmn}{d\eta_l\over\rho}{\partial_x\eta_m\over\rho}
{\partial_x\eta_n\over\rho}{\tau\over\rho}
\nonumber\\
&\ &\hspace{3.5cm}+24\Big[{dh\over \rho^2}+\Big({\eta\over\rho}\cdot
{d\eta\over\rho}\Big)\Big]
\epsilon_{lmn}{\partial_x\eta_l\over\rho}{\partial_x\eta_m\over\rho}{\partial_x\eta_n\over\rho}{\tau\over\rho}\Big\}
\nonumber\\
&\ &\hspace{2.5cm} -24\Big\{{dh\over\rho^2}\partial_x({\tau\over\rho}){\tau\over\rho}
+({\tau\over\rho}) d({\tau\over\rho})-\Big({\eta\over\rho}\cdot {d\eta\over\rho}\Big){\tau\over\rho}\partial_x ({\tau\over\rho})\Big\}. 
 \label{explicit-gamma}
\end{eqnarray}

\section{OSp(2$|$3)$_{target}$ and OSp(2$|$3)$_{diff}$} 
\setcounter{equation}{0}

The transformation (\ref{map}) defines  the finite form of the coadjoint action on $\widehat b + b$ along the orbit ${\cal O}_b$. Let it to be denoted  by 
\begin{eqnarray}
{\rm Ad}^*_{f,\varphi}b \equiv \Delta^{1\over 2}b(f,\varphi)+c{\cal S}(f,\varphi;x,\theta). 
\label{CCoad}
\end{eqnarray}
In this section we discover that 
it can have  OSp(2$|$3) symmetry  twofold. 
 One is the OSp(2$|$3) symmetry of the $N=3$ super-Schwarzian derivative ${\cal S}(f,\varphi;x,\theta)$ under the non-linear transformations by the Killing vectors (\ref{Killing}). It was proved in the end of Section 2.   We  show that it becomes   a symmetry of the 
  coadjoint action (\ref{CCoad}) as well when the initial point $b$  of  the coadjoint orbit ${\cal O}_b$ is chosen appropriately.   We call the symmetry {\it OSp(2$/$3)$_{target}$}.  
The other  is   a symmetry under the  superalgebra $\frak {osp}(2|3)$ given by  (\ref{superalgebra}),  which is a subalgebra of the  the $N=3$ superconformal algebra. 
It stabilizes the coadjoint action (\ref{CCoad}). That is,   
 we can show  that  the initial point $b$ of the orbit ${\cal O}_b$ is stable under the superalgebra $\frak {osp}(2|3)$, if $b$ is  tuned to be  a specific configuraton. Therefore  the coadjoint action (\ref{CCoad}) is invariant 
under  $\frak {osp}(2|3)$ as well.     
  We call this subsymmetry of the coadjoint action   {\it OSp(2$/$3)$_{diff}$}. 
 We shall discuss the two  OSp(2$|$3)  symmetries in detail.  

\subsection{OSp(2$|$3)$_{target}$}

It  seems that   the  the  coadjoint action (\ref{CCoad})  no longer   keeps the  OSp(2$|$3)$_{target}$ invariance 
 of the Schwarzian derivative,  due to the $b$-dependence. 
We show that  the  invariance remains if  the initial  point $b$ of the coadjoint orbit
${\cal O}_b$
is appropriately chosen. 
The idea is that 
for any  configuration of $b$ we may find 
 certain functions $F_0(x,\theta)$ and $\Phi_0(x,\theta)$ such that   
\begin{eqnarray}
 b(x,\theta) =  c{\cal S}(F_0,\Phi_0;x,\theta).
\label{delta-bS}
\end{eqnarray}
Here the Schwarzian derivative is regarded as a background configuration for $b(x,\theta)$. 
   The initial point $b(x,\theta)$ flows to $\Delta^{1\over 2}b(f,\varphi)$  along  the coadjoint orbit
    under the $N=3$ superconformal diffeomorphism. 
Putting (\ref{delta-bS}) into  (\ref{CCoad}) and using the composition law of the Schwarzian derivative (\ref{findiffeo}) we get  
\begin{eqnarray}
{\rm Ad}^*_{f,\varphi}b=
\Delta^{1\over 2}b(f,\varphi)+c{\cal S}(f,\varphi;x,\theta)
={\cal S}(F_0(f,\varphi),\Phi_0(f,\varphi);x,\theta). \label{renorm-S}
\end{eqnarray}
 We can say that the $b$-dependence of the coadjoint action (\ref{CCoad})  has been renormalized into the original Schwarzian derivative.  
Clearly this renormalized Schwarzian derivative  has  the same OSp(2$|$3)$_{target}$ invariance as   the original one because 
  the non-linear transformations (\ref{Killing})  flow as  
\begin{eqnarray}
\delta_\epsilon F_0 =\lim_{\mbox{\tiny$f\hspace{-0.05cm}=\hspace{-0.05cm}F_0, \varphi\hspace{-0.05cm}=\hspace{-0.05cm}\Phi_0$}}R^A(f,\varphi),\quad\quad
\delta_\epsilon \Phi_{0a}=\lim_{\mbox{\tiny$f\hspace{-0.05cm}=\hspace{-0.05cm}F_0, \varphi\hspace{-0.05cm}=\hspace{-0.05cm}\Phi_0$}}R_{\ a}^A(f,\varphi).
\label{renormalized}
\end{eqnarray}
Thus the coadjoint action (\ref{CCoad}) can keep OSp(2$|$3)$_{target}$ invariance.

The  background (\ref{delta-bS}) should obey the superconformal condition
\begin{eqnarray}
D_{\theta a}F_0=\Phi_{0b} D_{\theta a}\Phi_{0b}. \label{constraint}
\end{eqnarray}
Then it follows that  
\begin{eqnarray}
y={df+\varphi_c d\varphi_c\over \partial_x f+\varphi_c\partial_x\varphi_c}=
{dF_0(f,\varphi)+\Phi_{0c}(f,\varphi) d\Phi_{0c}(f,\varphi)\over \partial_x F_0(f,\varphi)+\Phi_{0c}(f,\varphi)\partial_x\Phi_{0c}(f,\varphi)}\equiv 
\lim_{\mbox{\tiny$f\hspace{-0.05cm}=\hspace{-0.05cm}F_0, \varphi\hspace{-0.05cm}=\hspace{-0.05cm}\Phi_0$}}
y.
\label{renorm-y}
\end{eqnarray}
This implies  that the one-form  $y$ in the Kirillov-Kostant two-form 
is invariant under the  renormalization of the $b$-dependence. Another important  property of  $y$ is that 
 it has also OSp(2$|$3)$_{target}$ invariance similarly to  the Schwarzian derivative. That is,   the denominator and the numerator respectively transform as 
\begin{eqnarray}
\delta_\epsilon(d f+\varphi_c d\varphi_c)&=&(\epsilon_{L^0}+2\epsilon_{\overline L}z+2\epsilon_{\overline F\hspace{0.05cm} a}\varphi_a)(d f+\varphi_c d\varphi_c), \nonumber\\
\delta_\epsilon(\partial_x f+\varphi_c\partial_x\varphi_c)&=&(\epsilon_{L^0}+2\epsilon_{\overline L}z+2\epsilon_{\overline F\hspace{0.05cm} a}\varphi_a)(\partial_x f+\varphi_c\partial_x\varphi_c),
\label{SchwarzInv'}
\end{eqnarray}
under the non-linear transformations (\ref{Killing}).
 
Owing to (\ref{renorm-S}) and (\ref{renorm-y})  the Kirillov-Kostant two-form $\widehat\Omega_b$ (\ref{KKorbit}) gets the $b$ dependence  renormalized.  
 The  renormalized  Kirillov-Kostant two-form   has  OSp(2$|$3)$_{target}$ invariance under the renormalized  non-linear transformations (\ref{renormalized}).

\subsection{OSp(2$|$3)$_{diff}$} 

The initial point $b(x,\theta) $ generically flows along the orbit ${\cal O}_b$ by the $N=3$ superconformal transformation. 
 But  it  could be stable  under  its subalgebra $\frak {osp}(2|3)$
  if $F_0$ and $\Phi_0$ in the initial configuration  (\ref{delta-bS}) are  constrained appropriately. 
 That is, we could have 
\begin{eqnarray}
 \lim_{f=x,\varphi=\theta}\Big(\delta_v{\rm Ad}^*_{f,\varphi}b(x,\theta)\Big)
&=& v\partial_xb(x,\theta)+{1\over 2}(\partial_xv)b(x,\theta)+{1\over 2}(D_{\theta b}v)D_{\theta b}b(x,\theta)+
cD_{\theta 1}D_{\theta 2}D_{\theta 3}v  \nonumber\\
&=& 0, \label{deltaAd'}
\end{eqnarray}
while $v$ is restricted to the submodes generated by  $\frak {osp}$(2$|$3) in (\ref{modeexp}). 
This stability implies  the 
 OSp(2$|$3)$_{diff}$ symmetry of the finite coadjoint action (\ref{CCoad}) which was  mentioned in the introduction of the section.  It  is different   from   OSp(2$|$3)$_{target}$ discussed previously.

We shall show that such an OSp(2$|$3)$_{diff}$ symmetry  indeed comes true by assuming  $b(x,\theta)$ to take  
 a specific expansion form  such that 
\begin{eqnarray}
b(x,\theta)=\theta_1\theta_2\theta_3d(x).
\label{b(x)}
\end{eqnarray}
 By expanding  $v$  as (\ref{v})
we find  (\ref{deltaAd'}) in components to be 
\begin{eqnarray}
 c\beta(x)+ \theta_a[c\partial_xt_a(x)]&+&{1\over2}\varepsilon_{abc}\theta_a\theta_b
[{1\over 4}\varepsilon_{abc}\beta_cd(x)+{c\over 2}\varepsilon_{abc}\partial_x^2\beta_c(x)]
\nonumber\\ 
 &+&\theta_a\theta_2\theta_3[ \alpha(x)\partial_x d(x)+2\partial_x\alpha(x) d+c\partial_x^3\alpha(x)]=0.  
\label{diff}
\end{eqnarray}
This gives as many  differential equations as the components of $v$. They can  be  solved by  
\begin{eqnarray}
\alpha(x)&=&e^{\pm inx}\alpha_{\pm},\quad \alpha_0,  \nonumber\\
 \beta_a(x)&=&e^{\pm {1\over 2}inx}\beta_{a{\hspace{0.05cm}}\pm{1\over 2}n},\nonumber\\
t_a(x)&=&t_{a\hspace{0.05cm}0}, \nonumber\\
 \beta(x)&=&0, 
\label{solution'}
\end{eqnarray}
when the configuration of $b$ is specified by 
\begin{eqnarray}
 d(x)={1\over 2}cn^2, 
\label{d(x)}
\end{eqnarray}
 with a fixed integer $n\in {\bf Z}_{\ne 0}$. With  $v$ restricted by (\ref{solution'})  the $N=3$ superconformal diffeomorphism  becomes  exactly  the superalgebra $\frak {osp}$(2$|$3)
  discussed in Section 2. Thus $b(x,\theta)$ given by (\ref{b(x)})  with  
(\ref{d(x)}) is a right  configuration for  the constraint (\ref{deltaAd'}) to be  satisfied.

 Putting (\ref{d(x)}) into (\ref{delta-bS})  we have  
\begin{eqnarray}
\theta_1\theta_2\theta_3\Big[{1\over 2}cn^2\Big]
=c{\cal S}(F_0,\Phi_0;x,\theta).
\nonumber
\end{eqnarray}
Finally we give a  solution of this equation  for  $F_0(x,\theta)$ and $\Phi_{0a}(x,\theta$).  
To this end  it is sufficient to know the non-supersymmetric part  of the r.h.s. of the equation, i.e., the non-supersymmetric Schwarzian derivative. 
 By using the expansion of the $N=3$ super-Schwarzian derivative in components  (\ref{Sexpansion}) and taking into account  the superconformal condition (\ref{constraint}) it turns out that 
\begin{eqnarray}
F_0(x,\theta)= {2\lambda^2\over  n} \tan({ nx\over 2}), \quad\quad \Phi_{0a}(x,\theta)=\theta_a\lambda\sec({ nx\over 2}),
\label{config}
\end{eqnarray}
with a constant $\lambda$.

With this solution for $F_0$ and $\Phi_0$ the finite coadjoint action (\ref{CCoad})
is invariant under OSp(2$|$3)$_{diff}$. 
On the other hand $y$ is also stable at the initial point of the orbit ${\cal O}_b$ as
\begin{eqnarray}
\lim_{f=x,\varphi=\theta}\delta_v y=0.
\label{limit-y}
\end{eqnarray}
 Therefore with the above  solution for $F_0$ and $\Phi_0$
the Kirillov-Kostant two-form  (\ref{KKorbit})   has OSp(2$|$3)$_{diff}$ invariance.

\vspace{0.5cm}

All the arguments about the twofold OSp(2$|$3) symmetry in this section can be straightforwardly applied to the cases of $N\le 2$ as well. But there is some difference  in  the application to the $N=4$ case. It is explained   in Appendix C.

\section{$N=3$ super-Schwarzian theory}

\setcounter{equation}{0}

In the end of Section 2 we have shown that 
 the Kirillov-Kostant two-form $\widehat\Omega_b$ on the coadjoint orbit
   satisfies
$i_v\widehat\Omega_b=dH$ with a zero-form.    
The $N=3$ super-Schwarzian theory may be defined by using this zero-form $H$ as 
\begin{eqnarray}
H|_{v=1}=\int dxd^3\theta\ {\rm Ad}^*_{f,\varphi}b\equiv \int dxd^3\theta\Big(\Delta^{1\over 2}b(f,\varphi)+c{\cal S}(f,\varphi;x,\theta)\Big).
\label{H}
\end{eqnarray}
The finite coadjoint action (\ref{CCoad}) is renormalized as  (\ref{renorm-S}) when $b$ is given by (\ref{delta-bS}). 
As shown in Subsection 4.1 it  has  OSp(2$|$3)$_{target}$ invariance  at any point of the renormalization flow. 
So does the action {(\ref{H}) being  written as 
\begin{eqnarray}
H|_{v=1}=c\int dxd^3\theta\ {\cal S}(F_0(f,\varphi),\Phi_0(f,\varphi);x,\theta).
\label{HH}
\end{eqnarray}

This action with $b\ne 0$ is not invariant under the superconformal diffeomorphism. This is  because the 
 renormalized Schwarzian derivative  transforms as (\ref{Schwarz2}). Note here that the anomalous part of the transformation disappears as  a boundary term in the  integration (\ref{HH}), but   the other part does not. 
When $b$ is further specified  as (\ref{b(x)}) with (\ref{d(x)}),  this configuration of $b$   is stable under OSp(2$|$3)$_{diff}$ satisfying the constraint (\ref{deltaAd'}). Therefore 
 the action (\ref{HH}) has the  OSp(2$|$3)$_{diff}$ symmetry as has been shown in Section 4.2. 


The action  (\ref{HH}) with this $b$ dependence deserves to be studied 
 by expanding $f$ and $\varphi$ in  components   as  in Appendix A.   
 We find that 
\begin{eqnarray}
H|_{v=1}&=&c\int dx\Big[
{n^2\over 2}\Big\{-{1\over 3}\epsilon_{ijk}\eta_i\eta_j\eta_k\tau {\partial_x\rho\over \rho}   
 +\rho^4-3\rho^2(\eta\cdot\partial_x\eta)+2(\eta\cdot\partial_x\eta)^2\Big\}\nonumber\\
 &\ & \hspace{1.2cm}+2\Big\{{1\over \rho}\partial^2_x\rho-2({\partial_x\rho\over\rho})^2
+{1\over \rho^2}\tau\partial_x\tau+{1\over \rho^2}(\partial_x\eta\cdot\partial_x^2\eta)\Big\}\Big],
\label{HHH}
\end{eqnarray}
with the constraints also given in Appendix A.   
  The first bracket is the top component of $\Delta^{1\over 2}\epsilon_{ijk}\varphi_i\varphi_j\varphi_k$, i.e., the $b$ dependent term with  (\ref{b(x)}), while the second bracket is the one of ${\cal S}(f,\varphi;x,\theta)$.  They are  exact results with no  approximation. The reader may refer to \cite{Note} for  details of the calculation.  To the quadratic  order of fermionic fields  the action turns out to take a fairly compact form     
\begin{eqnarray}
H_{quad}|_{v=1}&=& c\int dx\Big[{\cal S}(F_0,x)-{\cal S}(F_0,x)(\zeta\cdot\partial_x\zeta)
+2(\partial_x\zeta\cdot\partial_x^2\zeta)+{2\over \partial_x h}\tau\partial_x\tau   
\Big]. \label{quad}
\end{eqnarray} 
   Here we have used the constraint $\rho^2=\partial_xh+(\eta\cdot\partial_x\eta)=\partial_xh[1+(\zeta\cdot\partial_x\zeta)]$ with $ \eta_a=\sqrt {\partial_x h}\zeta_a$.  ${\cal S}(F_0;x)$ is the non-supersymmetric Schwarzian action  with $F_0={2\lambda^2\over  n}\tan{nh\over 2}$, given by (\ref{config}). It has resulted from  the use of  the composition  law (\ref{findiffeo}) at the non-supersymmetric limit. 
Now we are in  a position to evaluate this quadratic action at the stable point $h=x$. 
Then ${\cal S}(F_0,x)$  becomes  ${n^2\over 2}$, which  is the value at the stable point of the $N=3$ super-Schwarzian action (\ref{HH}). (\ref{quad}) becomes 
\begin{eqnarray}
H_{quad}|_{v=1}= c\int dx\Big[{n^2\over 2}-{n^2\over 2}(\zeta\cdot\partial_x
\zeta)+2(\partial_x\zeta\cdot\partial_x^2\zeta)+{2\over \partial_x h}\tau\partial_x\tau\Big]. 
\nonumber 
\end{eqnarray}

It is worth comparing this result with the one  for the case of $N=1$ or $N=2$, discussed in \cite{Super}. They can be obtained  by supersymmetry truncation $\tau=0$ and $(\eta\cdot\partial_x\eta)=\eta\partial_x\eta$ or $\eta_+\partial_x\eta_-+\eta_-\partial_x\eta_+$ for the respective case. 
 Firstly we comment on the $N=1$ case.  With the $b$ dependence  turned off the truncated action of (\ref{quad})  reduces to the 
  $N=1$ action which was given by (4.13)  in \cite{Super}.  
  To see this, set $n=0$ in (\ref{HH}). Then $F_0$  becomes $\lambda^2h$. As the result ${\cal S}(F_0;x)= {\cal S}(h;x)$. We find  the truncated action of  (\ref{quad}) to become  the $N=1$ action  given in \cite{Super}. Secondly we comment on the $N=2$ case. The truncated action of (\ref{quad}) is needed to be  modified by  a kind of U(1) anomaly. In \cite{Wi} they gave  an  interesting argument for its quantum effect. It is  due to chiral splitting  by the chirality condition.\footnote{The $N=2$ chiral superfields $\varphi_\pm$   are expanded in $\theta$ as 
\begin{eqnarray}
\varphi_\pm(x,\theta)&=&\eta_\pm(x)+\theta_\pm\rho(x)e^{\pm i\sigma(x) }+\theta_1\theta_2\partial_x\eta_\pm(x).  
\nonumber
\end{eqnarray}
They are independent of each other. }
  The chiral condition  allows us to U(1)-rotate  the fermionic components $\eta_{\pm}$  independently of each other.  Therefore supersymmetric truncation of the  action of  (\ref{quad})  should be done again by using   $\eta_\pm=\sqrt{\partial_xh} e^{\pm i\sigma} \zeta_\pm$ in place of  $\eta_\pm=\sqrt {\partial_x h}\zeta_\pm$. 
  Then there appear anomalous terms  with a phase factor  $\partial_x\sigma$. 
  We evaluate the action to be 
$$
H_{quad}|_{v=1}= c\int dx\Big[{n^2\over 2}-n^2\zeta_+\partial_x
\zeta_-+4\partial_x\zeta_+(\partial_x-iq)^2\zeta_-\Big], 
$$ 
 at the stable point where $h=x$ and $\partial_x\sigma=q=const$. It agrees with the 
 the result given by (3.48) in \cite{Wi}. 
On the contrary the $N=3$ theory has 
 O(3) symmetry, but no chirality condition. Therefore 
 although we may find  $\eta_\pm$, 
 in $\eta_a$ as O(2) transforming, they  do not have independent U(1) freedom from each other.  So there is no U(1) anomaly for the $N=3$ case. 
 A similar argument  is well-known  for chiral gauge theories, namely there is no chiral anomaly when the gauge group is real.

We may be interested in  quantizing  the $N=3$ theory given by (\ref{HH}).  In the final section we  comment on the Duistermaat-Heckman formula for the partition function discussed in \cite{Wi}.

\section{$N=(3,0)$ super-Liouville theory}

\setcounter{equation}{0}

In the end of Section 2 we have also shown that 
 the Kirillov-Kostant two-form  $\widehat\Omega_b$  on the coadjoint orbit
    can be written  in an  exact form   
$\widehat\Omega_b=dL$. Remember that $L$ is a one-form in the space of the cadjoint orbit.    
  The $N=(3,0)$ super-Liouville theory is given  by integrating $L$  along the orbit as 
\begin{eqnarray}
L=-\int_{{\cal O}_{b}} \int dxd^3\theta\Big\{y \Big(\Delta^{1\over 2}b(f,\varphi)+c{\cal S}(f,\varphi;x,\theta)\Big)
+{c\over 2}\gamma\Big\}. 
\label{I'}
\end{eqnarray}
The left-moving sector is described  by the $N=3$ superspace,  while  
  the right-moving sector by the extra space for the orbit  ${\cal O}_b$.  
It is worth checking the non-supersymmetric limit of the theory. The purely bosonic part of ${\cal S}(f,\varphi;x,\theta)$ has been already given by (\ref{Sexpansion}), while  the one of  $\gamma$  can be found in the first line in the formula (\ref{explicit-gamma}). $\Delta$ and $y$ take the familiar forms in the non-supersymmetric 
limit.  By the assumption (\ref{b(x)}) it follows that   
$$
\Delta^{1\over 2} b(f,\varphi)=\cdots +\rho\varphi_1\varphi_2\varphi_3 d(f)=\rho^4\theta_1\theta_2\theta_3,
d(h)+superpartners. 
$$  
Putting these results together and using the constraint (\ref{nonsusy-const}) we get 
\begin{eqnarray}
L= -\int_{{\cal O}_{b}}  \int dx{dh\over \partial_xh}\Big\{  (\partial_xh)^2d(h)+{c\over 2}\Big({\partial_x^3h\over\partial_x h}-2
({\partial_x^2h\over\partial_x h})^2\Big)+superpartners \ \Big\},
\nonumber
\end{eqnarray}
which is indeed  the the non-supersymmetric Liouville action given in \cite{Al}.

Let us study symmetries of the action. The $N=3$ superconformal invariance is evident  from the original form of the action (\ref{I'})  given by (\ref{LH-def})
\begin{eqnarray}
L=-\int dxd^3\theta (\widehat by+{c\over 2}\gamma).  \label{L-original}
\end{eqnarray}
 We may be interested in finding the energy-momentum tensor of the theory. The usual recipe for this is to recalculate $\delta_v L$  by assuming $dv\ne 0$. All the calculations in Section 3 have been done with $dv=0$. 
 With $dv\ne 0$ the formula  (\ref{deltay'}) for $\delta_vy$ is  modified  as
$$
 \delta_vy =dv+ v\partial_xy-(\partial_xv)y+{1\over 2}(D_{\theta c}v)D_{\theta c}y\equiv[v,y]={1\over 2}i_v[y,y].
$$
In calculating  $\delta_v\gamma$  the formula (\ref{delta-counter}) should be used in a  modified form
\begin{eqnarray}
\delta_v \int dxd^3\theta y D_{\theta 1}D_{\theta 2}D_{\theta 3}y&=& 
\int dxd^3\theta
(2dvD_{\theta 1}D_{\theta 2}D_{\theta 3}y) \nonumber\\
&+&
d\int dxd^3\theta
(2vD_{\theta 1}D_{\theta 2}D_{\theta 3}dy y).
\nonumber
\end{eqnarray}
Keeping in mind these contributions from  $dv \ne 0$  
  we find 
\begin{eqnarray}
\delta_v L=-\int_{{\cal O}_{b}}\int dxd^3\theta\ dv\Big(\Delta^{1\over 2}b(f,\varphi)+c{\cal S}(f,\varphi;x,\theta)\Big).
\nonumber
\end{eqnarray}
It may be written in the form 
\begin{eqnarray}
\delta_v I=\int_{{\cal O}_{b}}dt \int  dxd^4\theta\ v{d\over dt}\Big(\Delta^{1\over 2}b(f,\varphi)+c{\cal S}(f,\varphi;x,\theta)\Big),
\nonumber
\end{eqnarray}
by parametrizing the extra space of the  orbit with $t$. 
Therefore the  energy-momentum tensor of the theory  in the left-moving sector
 is given by the finite coadjoint action (\ref{CCoad}). 
 When $dv/dt=0$ it is conserved.  The appearance of the Schwarzian derivative  in energy-momentum tensor is the hallmark of the Liouville theory.  So far we have discussed the symmetry in the left-moving sector. It is obvious that the right-moving sector is  invariant under  reparametrization of the orbit.   
Thus the Liouville theory  given by (\ref{I'}) 
is invariant  under the $N=(3,0)$ superconformal diffeomorphism. It contrasts  with the Schwarzian theory given by  (\ref{H}).

When $b$ is given by  (\ref{delta-bS}), 
  the  Liouville action can be also rewritten in the renormalized form
\begin{eqnarray}
L&=&-\int_{{\cal O}_{b}} \int dxd^3\theta\ \Big\{cy{\cal S}(F_0(f,\varphi),\Phi_0(f,\varphi);x,\theta) +{c\over 2}{\gamma} \Big\}. \nonumber
\nonumber
\end{eqnarray}
As has been shown in Subsection 4.1 $y$ is invariant under the renormalization.  Therefore  the integrand of $L$ has  OSp(2$|$3)$_{target}$ invariance  at any point of the renormalization flow.

At this stage the action with (\ref{delta-bS}) is still invariant under the superconformal diffeomorphism on the contrary to the Schwarzian action. 
When $b$   is  further specified as (\ref{b(x)}) with (\ref{d(x)}), the action is no longer invariant under superconformal diffeomorphism. However 
we have  (\ref{limit-y}) as well as (\ref{deltaAd'}). Hence  $b$ is stable under OSp(2$|$3)$_{diff}$ and  the action 
  remains    invariant  under OSp(2$|$3)$_{diff}$. The reader may remember a  similar discussion   for the Kirillov-Kostant two-form in the end of Subsection 4.2.

\section{Conclusions}

The respective actions    (\ref{H}) and (\ref{I'})  are valid not only for the $N=3$ super-Schwarzian and Liouville theories, but also for the cases of $N\le 2$,   if ${\cal S}(f,\varphi;x,\theta)$,  $y$ and the anomalous term  therein are replaced   appropriately  for the  supersymmetries. For those quantities  see  Appendix A in \cite{A5}. 
We can argue the  actions for those cases similarly to  the $N=3$ case. 
  We obtain the same conclusions as in Sections 5 and 6. The twofold  symmetry is given by     OSp(2$|N$)$_{diff}$ and OSp(2$|N$)$_{target}$.  For the $N=4$ case the similar  arguments 
 go through and end up with almost the same conclusions. But there is a difference.  It  will be argued in Appendix C.

 On top of the feature summarized above  there is one  more point worthy to be remarked.  For the Liouville theories with $N\le  2$   
the OSp(2$|N$)$_{target}$ symmetry  becomes  local in the right-moving sector. For those theories 
  the anomalous terms can be  expressed in   exact forms  in terms of superfields. They are given by 
\begin{eqnarray}
\int dx y\partial_x^3y &=& -d\int dx \Big[y\Big({\cal S}+{1\over 2}({\partial^2 h\over \partial h})^2\Big)\Big],  \nonumber\\
    \nonumber \\
\int dxd\theta{1\over 2}yD_\theta\partial_x^2y &=&-d\int dxd\theta \Big[y\Big({\cal S}+\Big({D_\theta^3\varphi\over D_\theta\varphi}{D_\theta^2\varphi\over D_\theta\varphi}\Big)\Big)\Big],  
    \nonumber\\
    \nonumber \\
\int dxd^2\theta  {1\over 2}y\partial_x[D_{\theta+},D_{\theta-}]y&=& -d\int dxd^2\theta \Big[y\Big({\cal S}-2{\partial_x\varphi^+\over D_{\theta +}\varphi^+}{\partial_x\varphi^-\over D_{\theta -}\varphi^-}  \Big)\Big],  \nonumber
\end{eqnarray} 
 for $N=(0,0),(1,0),(2,0)$ respectively. 
Owing to  these formulae we can show that the  respective Liouville actions  are invariant locally in the right-moving sector
under  non-linear transformations  by  the Killing vectors  like (\ref{Killing}).
 That is, they are invariant even  when the  infinitesimal parameters    are local as  $\epsilon^A(t)$\cite{Al,A2,A3}.  
On the contrary    the anomalous term in  (\ref{L-original})  for the case of $N=3$ did  not admit such a supercovariant  expression. 
 It did not  for  the case of $N=4$ either. The anomalous term $\gamma$ in (\ref{L-original}) 
 can get a local expression 
   only  in terms of components as (\ref{explicit-gamma}). 
 For the $N=4$ case the reader may refer to \cite{A5}. Without a supercovariant expression of $\gamma$   we can hardly study local invariance under   OSp(2$|N$)$_{target}$ or 
 PSU(1,1$|$2)$_{target}$  in the right-moving sector. 
 
 In this paper  the arguments  have been limited at the classical level.
 Obtaining a simple form of the action (\ref{quad}) 
one may be interested in a saddle-point calculation of the partition function of the $N=3$ super-Schwarzian theory. One may ask about one-loop exactness of the calculation.  Here  the measure of the path-integration does matter.   In \cite{Wi} it was worked out for  the non-supersymmetric  case and one-loop exactness of the partition function was shown   by means of  the Duistermaat-Heckman formula. It was also pointed out  that  such an argument  is applicable in supersymmetric cases as long as it is formulated  by the coadjoint orbit method.  We have all the materials at hand to discuss  the issue for the $N=3$ case as well.  In Section 4 we have discussed OSp(2$|$3)$_{diff}$ invariance of the Kirillov-Kostant 2-form $\widehat\Omega_b$ as a residual symmetry after the gauge-fixing.  Hence the relevant symplectic supermanifold is $diff(S^{1|3})/{\rm OSp}(2|3)$. 
It is described by the component fields $f,\eta_1,\eta_2,\eta_3$ of the $N=3$ superconformal symmetry and the ones with opposite grading, which are denoted by  $df,d\eta_1,d\eta_2, d\eta_3$ according to \cite{Wi}. (Here $d$  no longer means an exterior derivative, but it is used as a convention to denote partner fields avoiding  new naming.) 
 The Duistermaat-Heckman formula for the partition function  reads  as 
\begin{eqnarray}
Z=\int{d\mu(f,\eta_1,\eta_2,\eta_3,df,d\eta_1,d\eta_2, d\eta_3)\over {\rm Osp}(2|3)} \exp[\widehat\Omega_b +H|_{v=1}],  \label{Z}
\end{eqnarray}
with the Kirillov-Kostant 2-form and the $N=3$ super-Schwarzian action given by given by (\ref{KKorbit}) and (\ref{HHH}) respectively. 
However we may immediately note that both actions in the exponent  contain the O(3)-singlet fermionic field $\tau$ and its odd-graded partner $d\tau$. We then ask how to path-integrate them. 
 $\tau$   is related to the generator $F(x)$ in (\ref{T})  of the  $N=3$ superconformal algebra discussed  in ii) of Section 2. 
 As known in the literature this generator 
 may be  eliminated from the superconformal algebra to get  the algebra in a non-Lie  algebraic form\cite{Kniz,Ber}. Correspondingly we think of eliminating the dependence on $\tau$ and $d\tau$ from the action $\widehat\Omega_b +H|_{v=1}$
 in favour of      $\eta$ and $d\eta$. But it can be hardly done  even by using   the constraints in Appendix A. $\tau$ and $d\tau$ are not constituents of 
  the symplectic supermanifold $diff(S^{1|3})/{\rm Osp}(2|3)$. 
In the presence of them it is a point at issue whether  the Duistermaat-Heckman formula is still applicable for the $N=3$ super-Schwarzian theory. The partition function might loose the reason for  one-loop exactness for the $N=3$ theory. The author will study the issue furthermore.

\appendix

\section{The superconformal condition}
\setcounter{equation}{0}

From the superconformal condition (\ref{cond}) it follows that  
\begin{eqnarray}
D_{\theta a}\varphi_cD_{\theta b}\varphi_c=D_{\theta c}\varphi_aD_{\theta c}\varphi_b=
\delta_{ab}(\partial_x f+\varphi_c\partial_x\varphi_c).
\label{useful}
\end{eqnarray}
Hence the scaling factor in (\ref{findiffeo}) can be written as 
\begin{eqnarray}
\Delta = \partial_x f+\varphi_c\partial_x\varphi_c.
\nonumber
\end{eqnarray}
The similar relations can be found also for the case of $N=1,2$ and 4 superconformal diffeomorphisms. Keep in mind  that $\Delta$ is a superconformal field with weight 1.  

We can write the superconformal condition (\ref{cond})  in components. $f$ and $\theta_a$ are  expanded  in $\theta$ as  
\begin{eqnarray}
f(x,\theta)&=&h(x)+\theta_c\psi_c(x)+{1\over 2}\varepsilon_{abc}\theta_a\theta_bt_c(x)+\theta_a\theta_2\theta_3\omega(x), \nonumber\\
\varphi_a(x,\theta)&=&\eta_a(x)+\theta_a\rho(x)+{1\over 2}\varepsilon_{abc}\theta_b\theta_c\tau(x)+\theta_a\theta_c\tau_c(c)
+\theta_1\theta_2\theta_3r_a(x).
\label{expansion}
\end{eqnarray}
Putting these into the superconformal condition (\ref{cond})  gives  the constraints 
\begin{eqnarray}
\partial_x h&=& -\eta_c\partial_x\eta_c+\rho^2,\quad\quad \psi_a=\eta_a\rho, \nonumber\\
t_a&=&-\eta_a\tau-\varepsilon_{abc}\eta_b\tau_c,\quad\quad \omega=\eta_c\tau_c-\tau\rho, \nonumber\\
\tau_a&=&\partial_x\eta_a,\quad\quad 0=\rho r_a+\tau\tau_a+{1\over 2} 
\varepsilon_{abc}\tau_b\tau_c.
\nonumber
\end{eqnarray}

\section{Proofs of (\ref{Omega-closure}) and (\ref{delta-counter})}
\setcounter{equation}{0}

 The integrand of  (\ref{Omega-closure})  can be rewritten as 
\begin{eqnarray}
3\epsilon_{abc}y(D_{\theta a}y)D_{\theta b}D_{\theta c}\partial_x y=-\epsilon_{abc}(y\partial_xy )D_{\theta a}D_{\theta b}D_{\theta c} y +boundary\ terms. 
\label{A0}
\end{eqnarray}
We shall show the formula in this form. 
The l.h.s. can be calculated  in two ways
\begin{eqnarray}
\epsilon_{abc}y(D_{\theta a}y)D_{\theta b}D_{\theta c}\partial_x y
&=&A+D_{\theta a}(\cdots) \label{A1},\\
\epsilon_{abc}y(D_{\theta a}y)D_{\theta b}D_{\theta c}\partial_x y 
&=& -B-A  +\partial_x(\cdots), \label{A2}
\end{eqnarray} 
in which
\begin{eqnarray}
A&=&\epsilon_{abc}y(D_{\theta a}\partial_xy)D_{\theta b}D_{\theta c} y, \nonumber\\
B&=&\epsilon_{abc}\partial_xy(D_{\theta a}y)D_{\theta b}D_{\theta c} y.\nonumber
\end{eqnarray} 
(\ref{A2}) can be put in one other form by calculating the r.h.s. as 
\begin{eqnarray}
\epsilon_{abc}y(D_{\theta a}y)D_{\theta b}D_{\theta c}\partial_x y
&=& -2A-\epsilon_{abc}(y\partial_xy)D_{\theta a}D_{\theta b}D_{\theta c} y +D_{\theta a}(\cdots). 
\label{A3}
\end{eqnarray} 
The three equations can not be  independent. Eliminating $A$ and $B$ we find (\ref{A0}).

Next we show (\ref{delta-counter}), i.e., 
\begin{eqnarray}
\delta_v\int dxd^3\theta\epsilon_{abc}yD_{\theta a}D_{\theta b}D_{\theta c}y&=&
2d\int dxd^3\theta\epsilon_{abc} vD_{\theta a}D_{\theta b}D_{\theta c}y.
\label{d2-form}
\end{eqnarray}
By using (\ref{dy'}) and (\ref{deltay'}) we calculate both sides of the equation as 
\begin{eqnarray}
 {\rm l.h.s.}&=&2\int dxd^3\theta\Big( \epsilon_{abc} [v,y]D_{\theta a}D_{\theta b}D_{\theta c}y\Big)\nonumber\\
&=&2\int dxd^3\theta\epsilon_{abc}\Big(
 (-{1\over 2}v\partial_xy-(\partial_xv)y)D_{\theta 1}D_{\theta 2}D_{\theta 3}y
+{3\over 2}v(D_{\theta a}y)D_{\theta b}D_{\theta c}\partial_x y\Big), \nonumber\\
{\rm r.h.s.}&=&2\int dxd^3\theta\epsilon_{abc} vD_{\theta a}D_{\theta b}D_{\theta c}d y
=-2\int dxd^3\theta \epsilon_{abc}dyD_{\theta a}D_{\theta b}D_{\theta c}v
\nonumber\\
&=&-2\int dxd^3\theta\epsilon_{abc}(y\partial_xy+{1\over 4}D_{\theta d}yD_{\theta d}y)D_{\theta a}D_{\theta b}D_{\theta c}v.
\nonumber
\end{eqnarray}
By repeating  integration by parts in  the r.h.s. the integrand becomes 
\begin{eqnarray}
&\epsilon_{abc}&\Big(\hspace{-0.1cm}-yD_{\theta a}\partial_xy+{1\over 2}(D_{\theta a}y)\partial_xy\Big)
D_{\theta b}D_{\theta c}v+D_{\theta a}(\cdots) \nonumber\\
=&\epsilon_{abc}&\Big(\hspace{-0.1cm}-{1\over 2}(D_{\theta b}y)D_{\theta a}\partial_xy-yD_{\theta b}D_{\theta a}\partial_xy+{1\over 2}(D_{\theta b}D_{\theta a}y)\partial_xy\Big)D_{\theta c}v+D_{\theta b}(\cdots)\nonumber\\
=&\epsilon_{abc}&\Big(\hspace{-0.1cm}-{3\over 2}(D_{\theta b}y)D_{\theta c}D_{\theta a}\partial_xy+yD_{\theta c}D_{\theta b}D_{\theta a}\partial_xy+{1\over 2}(\partial_xy)D_{\theta c}D_{\theta b}D_{\theta a}y\Big)v+D_{\theta c}(\cdots).
\nonumber
\end{eqnarray}
Integrating this over the superspace  we   find the l.h.s.

\section{The $N=4$ super-Schwarzian theory revisited}

\setcounter{equation}{0}

The  issue  for the $N=3$ Schwarzian theory in Section 5 were studied also for the $N=4$ super-Schwarzian action in \cite{AoHo}. However  the $b$ field renormalization was not discussed. In this Appendix we complete   the arguments for $b\ne 0$ giving the missing part. 
We show that the $N=4$ super-Schwarzian action  can also have  the two symmetries similarly to the $N=3$ action. They are PSU(1,1$|$2)$_{target}$ and PSU(1,1$|$2)$_{diff}$, but  do not  appear simultaneously. It contrasts with the $N\le 3$ case.  
Moreover we show that they are not symmetries of the action density. This feature  also makes the  $N=4$ super-Schwarzian action  different from the $N\le 3$ case. 

To explain these features  let us remember  
the $N=4$ super-Schwarzian action, which corresponds  to (\ref{H}), i.e.,
\begin{eqnarray}
H=\int dxd^4\theta\ {\rm Ad}^*_{f,\varphi}b=\int dxd^4\theta \Big( b(f,\varphi)+c{\cal S}(f,\varphi;x,\theta)\Big).
\label{N=4action}
\end{eqnarray}
(See (5.14) in \cite{AoHo}.)   When $b$ is given by 
 $
 b(x,\theta) =  c{\cal S}(F_0,\Phi_0;x,\theta),
$
 the finite coadjoint action ${\rm Ad}^*_{f,\varphi}b$ is renormalized  similarly to the $N=3$ case, i.e., 
\begin{eqnarray}
{\rm Ad}^*_{f,\varphi}b(x,\theta)  &=& b(f,\varphi)+c{\cal S}(f,\varphi;x,\theta)
    \nonumber \\
&=& c{\cal S}(F_0(f,\varphi),\Phi_0(f,\varphi);x,\theta).
\label{ReN4}
\end{eqnarray} 
 The action $H$  has    PSU(1,1$|$2)$_{target}$ symmetry at any point of the renormalization  flow.  But it is not a symmetry of the action density. It is because 
 the $N=4$ super-Schwarzian derivative ${\cal S}(f,\varphi;x,\theta)$ transforms   as (\ref{N4})  under non-linear transformations realized by 
the Killing vectors like (\ref{Killing}). 
 In \cite{AoHo} it was shown that the breaking terms (\ref{N4})  disappear by the integration in (\ref{N=4action}) as boundary terms.

So far the action $H$ with 
 $
 b(x,\theta) =  c{\cal S}(F_0,\Phi_0;x,\theta)
$
is not invariant under the superconformal diffeomorphism, since the integrand transforms as a superconformal field with $w=0$. Note here that the anomalous part of the transformation disappears as  a boundary term in the  integration (\ref{N=4action}), 
but the other part does not. This is also a feature we have seen for the $N=3$ super-Schwarzian theory.

Now  we may ask whether 
the action  (\ref{N=4action})  can have  the PSU(1,1$|$2)$_{diff}$ symmetry 
 as the $N=3$ theory had the OSp(2$|$3)$_{diff}$ symmetry. The  question is whether we could have  for the action (\ref{N=4action})  
\begin{eqnarray}
  \delta_v H&=&\int dx d^4\theta \lim_{f=x,\varphi=\theta}\Big(\delta_v{\rm Ad}^*(f,\varphi)b(x,\theta)\Big)\nonumber\\
 &=& \int dx d^4\theta\Big\{v\partial_xb(x,\theta)+{1\over 2}D_{\theta a}vD_{\theta}^{\ a}b(x,\theta)+{1\over 2}D_{\theta}^{\ a}vD_{\theta a}
b(x,\theta)+
c\partial_xv\Big\}, \hspace{1cm}\nonumber\\
 &=& 0
\label{deltaAd2}
\end{eqnarray} 
while restricting  the parameter $v$ to the modes of the subalgebra $\frak {usp}$(1,1$|$2). 
 It was shown in \cite{AoHo} that this constraint  is satisfied when $b(x,\theta)$ takes 
 the  specific configuration  
\begin{eqnarray}
b(x,\theta) = (\theta_a\theta^a)^2(-{1\over 4}cn^2) + superpartners, 
\label{N=4,b}
\end{eqnarray}
given by (5.19)  therein.   
Therefore with  this configuration of $b$  the action has  the PSU(1,1$|$2)$_{diff}$ symmetry. But we then loose the PSU(1,1$|$2)$_{target}$ symmetry. This is because with the configuration (\ref{N=4,b}) we can not find functions $F_0(x,\theta)$ and $\Phi_0(x,\theta)$
satisfying 
$
 b(x,\theta) =  c{\cal S}(F_0,\Phi_0;x,\theta)
$.  
For this it is enough to note that the $N=4$ super-Schwarzian derivative takes the form in the non-supersymmetric limit
\begin{eqnarray}
{\cal S} (f,\varphi;x,\theta)
&=& \log \partial_xh+{1\over 2}(\theta_a\theta^a)^2\Big[-{\partial_x^3h\over \partial_xh}+2({\partial_x^2h\over \partial_xh})^2\Big]+superpartners.
\label{approx}
\end{eqnarray} 
(See (3.17) in \cite{AoHo}.)   
The term $\log   \partial_x h$ hinders  from equating (\ref{N=4,b}) and (\ref{approx}). 
 Its appearance   is consistent with the transformation law (\ref{N4}). 
  It  is characteristic for the $N=4$ super-Schwarzian derivative in contrast with the case of $N\le 3$. 

Even though we have lost the PSU(1,1$|$2)$_{target}$ symmetry we may be interested in 
the PSU(1,1$|$2)$_{diff}$ symmetry by itself. We then  wonder if it might be a symmetry of  the action  density (\ref{ReN4}). 
As shown in \cite{AoHo} it is too strong to require the integrand in (\ref{deltaAd2}) to be vanishing while restricting  $v$ to the modes of the subalgebra $\frak {usp}$(1,1$|$2). 
There is no way to gauge-fix the $b$ field so that it is stable at the initial point of the coadjoint orbit under  $\frak {usp}$(1,1$|$2).  This point is also different from the $N\le 3$ case.  


\end{document}